\documentclass[10pt,sigconf,nonacm,letterpaper]{acmart}

\usepackage{booktabs} 
\usepackage{soul}
\usepackage{color}
\usepackage[inline]{enumitem}
\usepackage{multirow}
\usepackage{listings}
\usepackage[normalem]{ulem}
\usepackage{subcaption}
\usepackage{color, colortbl}
\definecolor{Gray}{gray}{0.9}

\setcopyright{acmcopyright}
\renewcommand\footnotetextcopyrightpermission[1]{} 
\acmConference{IMC '20}{October 27--29, 2020}{Pittsburgh, USA}

\author{Jacky Cao}
\affiliation{%
 \institution{University of Oulu}
 \city{Oulu}
 \country{Finland}
} 
\email{jacky.cao@oulu.fi}

\author{Xiang Su}
\affiliation{%
 \institution{University of Helsinki}
 \city{Helsinki}
 \country{Finland}
}
\affiliation{%
 \institution{ University of Oulu}
 \city{Oulu}
 \country{Finland}
} 
\email{xiang.su@helsinki.fi}

\author{Benjamin Finley}
\affiliation{%
 \institution{University of Helsinki}
 \city{Helsinki}
 \country{Finland}
}
\email{benjamin.finley@helsinki.fi}

\author{Pengyuan Zhou}
\affiliation{%
 \institution{University of Helsinki}
 \city{Helsinki}
 \country{Finland}
}
\email{pengyuan.zhou@helsinki.fi}

\author{Pan Hui}
\affiliation{%
 \institution{University of Helsinki}
 \city{Helsinki}
 \country{Finland}
}
\affiliation{%
 \institution{The Hong Kong University of Science and Technology}
 \city{Hong Kong}
}
\email{pan.hui@helsinki.fi}

\begin{document}
\title{Evaluating Transport Protocols on 5G for Mobile Augmented Reality}

\begin{abstract}
Mobile Augmented Reality~(MAR) mixes physical environments with user-interactive virtual annotations. Immersive MAR experiences are supported by computation-intensive tasks which rely on offloading mechanisms to ease device workloads. However, this introduces additional network traffic which in turn influences the motion-to-photon latency (a determinant of user-perceived quality of experience). Therefore, a proper transport protocol is crucial to minimise transmission latency and ensure sufficient throughput to support MAR performance. Relatedly, 5G, a potential MAR supporting technology, is widely believed to be smarter, faster, and more efficient than its predecessors. However, the suitability and performance of existing transport protocols in MAR in the 5G context has not been explored. Therefore, we present an evaluation of popular transport protocols, including UDP, TCP, MPEG-TS, RTP, and QUIC, with a MAR system on a real-world 5G testbed. We also compare with their 5G performance with LTE and WiFi. Our evaluation results indicate that TCP has the lowest round-trip-time on 5G, with a median of $15.09\pm0.26$ ms, while QUIC appears to perform better on LTE. Through an additional test with varying signal quality (specifically, degrading secondary synchronisation signal reference signal received quality), we discover that protocol performance appears to be significantly impacted by signal quality.
\end{abstract}


\maketitle
\thispagestyle{empty}

\section{Introduction}
Mobile augmented reality~(MAR) supplies users with additional information and enhanced perception of their surroundings through superimposed virtual augmentations on real-time camera feeds~\cite{azuma2001recent, hollerer2004mobile}. The camera image frame is thus the prevalent data source for MAR applications and processing of these images demands high computational capacity and therefore impacts device usability and battery life. Offloading addresses this challenge by leveraging the computation and storage capabilities of external servers. Researchers have proposed cloud and edge offloading-based systems to improve performance and minimise delay. While cloud computing provides powerful centralised remote resources, edge computing places resources close to the end users~\cite{chang2014bringing}, resulting in lower client-server latencies~\cite{shi2016edge} and fewer bottlenecks~\cite{varghese2016challenges}. Furthermore, edge computing enables better security and privacy protection, e.g., by limiting the effect of distributed denial of service attacks and allowing for location-based authentication \cite{zhang2018data}. To leverage the benefits of edge computing, a suitable network transport protocol is required.

However, no existing transport protocol is designed and optimised for MAR. We envision that an ideal tailored MAR multimedia transport protocol should
\begin{enumerate*}
    \item handle transmission of various data types under varying requirements (latency, jitter, bandwidth, and integrity) and priorities;
    \item achieve a good balance between fairness and exploit the maximum bandwidth;
    \item provide low latency and high fault tolerance to allow real-time communication;
    and \item support multipath transmission~\cite{braud2017future}.
\end{enumerate*}

Unfortunately, given the above considerations, more research is required on the suitability of existing protocols in MAR, especially in the context of 5G. Therefore, this work focuses on the evaluation of diverse transport protocols in a MAR system on a 5G testbed. We present these 5G evaluation results along with LTE and WiFi results as two comparative baselines. We evaluate several selected protocols, specifically UDP, TCP, MPEG-TS, RTP, and QUIC, according to key network metrics including round-trip time~(RTT), jitter, and throughput.

Our 5G evaluation results suggest that TCP is the most suitable protocol for MAR applications on 5G. TCP shows the lowest RTT values for seven out of nine tested image resolutions, with a median of $15.09\pm0.26$ ms. By comparison, QUIC performs better with LTE connections. Additionally, through an evaluation with varying signal quality (specifically, different SS-RSRQs\footnote{Degrading secondary synchronisation signal reference signal received quality}), we find that signal quality appears to significantly impact protocol performance, especially QUIC.

Our work is one of the first efforts to evaluate transport protocols for MAR on 5G. We make the following contributions when compared to previous work~(Section \ref{ssec:related_work}):
\begin{itemize}
    \item \textit{Comparison of diverse transport protocols in a MAR system.} We provide a quantitative performance evaluation of five transport protocols, including general transport protocols such as UDP, TCP, and QUIC and multimedia intended transport protocols, such as RTP and MPEG-TS.
    \item \textit{Evaluation with a real-world 5G testbed.} We conduct our evaluation with a real-world 5G testbed (including 5G smartphone, 5G base station, and edge server) and compare the performance of 5G, LTE and WiFi. 
    \item \textit{Performance analysis through various metrics.} We perform a comprehensive analysis focusing on networking metrics.
\end{itemize}

The remainder of the paper is organised as follows. Section \ref{sec:background} presents an overview of the transport protocols and related work in the area of transport protocol analysis. Section \ref{sec:methods} details the evaluation setup, methods, and metrics. Section \ref{sec:analysis} presents the evaluation of the collected network measurement results and Section \ref{sec:conclusions} concludes the paper.

\section{Background and Related Work} \label{sec:background}
We briefly introduce 5G, MAR, and requirements of MAR from a networking perspective. Then, we present the transport protocols we evaluate and summarise related work.

\subsection{5G and MAR}
5G is the new generation wireless communication standard enabling a paradigm shift that includes very high carrier frequencies with massive bandwidths, extreme base station and device densities, and unprecedented numbers of antennas. The key features of 5G include ultra-low latency, increased data transmission rates, massive multiple-input multiple-output capabilities, network slicing, and the ability to accommodate a much larger set of devices \cite{agiwal2016next, foukas2017network}. Current LTE and WiFi technologies do not match the proposed capabilities of 5G. The increased capacity and shorter latency introduced by 5G may influence protocol performance.

Because MAR is highly latency-sensitive, 5G benefits MAR with reduced latencies to offload computation to external servers. Additionally, extensive computation offloading requires substantial network bandwidth provided by 5G.
Hu \textit{et al.} \cite{hu2016quantifying} quantifies the impact of edge computing on mobile applications using both LTE and WiFi connections. They conclude that edge computing improves response time and lowers device energy consumption, which is especially notable on LTE. 5G is believed to further reduce end-to-end latency and support immersive experiences for MAR.

\subsection{Transport Protocols}
\textbf{User Datagram Protocol (UDP)}
UDP \cite{postel1980user} utilises a datagram mode of communication between network hosts and over the Internet Protocol (IP). Data is transmitted between applications with a transaction orientated protocol mechanism, so delivery order and protection against duplicate data is not guaranteed.

\textbf{Transmission Control Protocol (TCP)}
TCP \cite{postel1981transmission} is designed to be a highly reliable protocol built on IP for communication between network hosts. TCP ensures that data is recoverable in the event of issues (e.g., data loss, corruption, or out-of-order delivery) and utilises packet sequence numbering and acknowledgement packets to provide this reliability.

\textbf{MPEG Transport Stream (MPEG-TS)}
MPEG-TS is a purpose built protocol for transporting multimedia content. MPEG-TS uses transport streams, wherein data streams are multiplexed into a single signal and decoded into individual channels when received \cite{SCHONFELD20051031}. MPEG-TS is used to transmit real-time video and audio in traditional television applications \cite{HASKELL200289, 2012555} and in Internet scenarios where MPEG-TS is run over UDP, i.e., MPEG-4 encoded videos are placed into UDP packets \cite{FARMER2017273}.  

\textbf{Real-time Transport Protocol (RTP)}
RTP \cite{ietf2003rtp} is a protocol designed to transfer real-time audio and video between network end hosts. RTP streams are monitored by a feedback mechanism, the real-time control protocol (RTCP). However, RTP does not manage resource reservation or guarantee QoS. RTP is typically run on top of UDP, but is also able to function over other transport protocols, e.g., TCP \cite{PORTER200745}.

\textbf{Quick UDP Internet Connections (QUIC)}
QUIC \cite{bishop2019hypertext} is a transport protocol aiming to provide low latency and secure data transport over UDP. QUIC is built with HTTP/3 and provides multiplexing and flow control, which uses TLS equivalent security measures and TCP-like congestion control. QUIC packets are authenticated and associated payloads encrypted, and lost packets are recovered through forward error correction.

\subsection{Evaluation of Transport Protocols} \label{ssec:related_work}
There are limited prior works exploring the performances of transport protocols in MAR scenarios. Therefore, the most related works are the comparisons of transport protocols in the context of other networks and applications.

Akan \cite{akan2007performance} comprehensively evaluates transport protocol performance for multimedia in Wireless Sensor Networks (WSN) and concludes that new transport protocols are needed to satisfy the requirements of WSN multimedia transmission. 

For transport protocol evaluation in traditional IP networks, Camarillo \textit{et al.} \cite{camarillo2003evaluation} evaluates SCTP for signalling transport with a focus on the mean delay metric. Their simulations show that there is no substantial performance increase for SCTP over TCP for head of the line (HOL) blocking in normal conditions (without large traffic loss), and for better SCTP performance, larger HOL blocking is required in simulations. Johnsen \textit{et al.} \cite{johnsen2013evaluation} assesses four different protocols (UDP, SCTP, TCP, and AMQP) for web traffic in military networks with the metric of success rate of data transfer. They confirm that UDP does not provide guaranteed end-to-end delivery, that SCTP generates more overhead than TCP given a lower bandwidth connection, and that AMQP produces a considerable amount of signalling traffic between client and broker which leads to low success rates at low bandwidths. 

Furthermore, as a promising newer protocol, QUIC has been extensively evaluated. Megyesi \textit{et al.} \cite{megyesi2016quick} evaluate QUIC, SDP, and HTTP based on metrics such as page load time and packet loss. They conclude that the best protocol depends on network conditions. Biswal \textit{et al.} \cite{bishop2019hypertext} perform a similar comparison with HTTP/2 versus QUIC+SPDY 3.1 while utilising 2G and LTE connections and conclude that $90\%$ of pages load faster with QUIC over 2G and $60\%$ of pages over LTE.

\section{Methods and Data Collection} \label{sec:methods}
\subsection{MAR System on 5G Testbed}
\begin{figure}
    \begin{center}
      \centering
      \includegraphics[width=\linewidth]{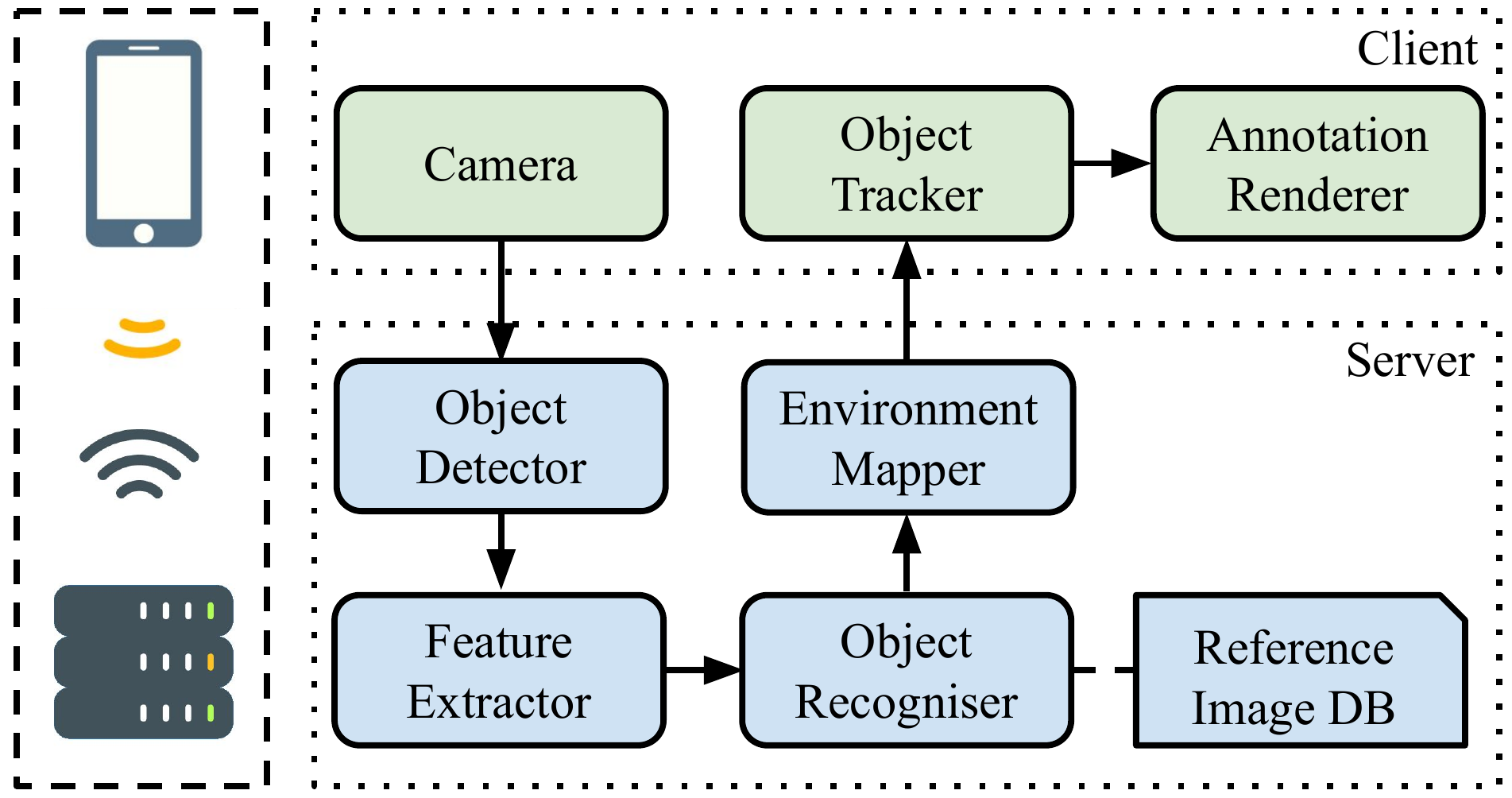}
      \caption{Pipeline of the MAR system.}
      \label{fig:mar_system}
    \end{center}
\end{figure}
Our MAR system is composed of an Android client application which captures images and generates object recognition requests and a server application which performs object recognition and sends the results back to the client. The system pipeline has the following steps (Figure \ref{fig:mar_system}): 1) the client captures images from the device camera at a rate of one image every 33 ms; 2) the images are sent to the server at the same rate and the server performs object detection; 3) if objects are detected, their features are extracted; 4) extracted feature points are analysed by the object recogniser which retrieves the object information from a reference image database; 5) if there are recognised objects, the environment mapper calculates the object bounding boxes; 6) the box vertices are sent to the client object tracker; and 7) the annotation renderer draws the virtual information on the client display.

In our MAR system, data is transferred between client and server during two distinct steps: 1) when the client sends an object recognition request to the server and 2) when the server replies with the request results. For both transfers, the required information is structured as a byte array and subsequently passed to the network transport protocols.

The size of the request byte array depends on the request image data size (which is a function of the resolution, the compression method, and the image entropy). The image (directly from the camera) is first pre-processed before being placed in the request array. Specifically, the image is 1) resized to a standard resolution of $4000\times1824$ pixels, 2) grayscaled, and 3) then down-scaled to a smaller resolution to reduce packet size.


The client application runs on a OnePlus 7 Pro 5G Android smartphone. While the server application runs on an edge PC with an Intel Core i7-9750H CPU, 32 GB memory, and an NVIDIA GeForce RTX 2080 Max-Q GPU and is connected to the 5G testbed through an Ethernet connection. The 5G testbed contains a base transceiver station (BTS) and two antennae operating in the C-Band spectrum for 5G at 3500 MHz with a bandwidth of 60 MHz, additionally, LTE is provided by an LTE Picocell at 2600 MHz with a bandwidth of 10 MHz. The distance from the outdoor 5G BTS to the indoor smartphone is approximately 30~m with no direct line of sight (NLOS), while the distance from the indoor LTE Picocell is approximately 10m also with NLOS. Finally, the indoor edge PC generates a 5 GHz 802.11ac WiFi hotspot and the distance from the PC is approximately one meter with line of sight (LOS). We argue that these are typical conditions, e.g., mobile networks access points (APs) are less often in direct LOS with clients, especially indoors. However, we note that the studied conditions and contexts are a starting point and future work should expand on the variety of conditions analysed.


\subsection{Evaluation and Performance Metrics}
\begin{table}
  \caption{Summary of evaluation parameters}
  \label{tab:experiment_parameters}
  \begin{tabular}{cc}
    \toprule
    Parameter & Values \\
    \midrule
    Protocol & UDP, TCP, MPEG-TS, RTP, QUIC \\
    Image resolution & $1152\times648$, $1024\times576$, $896\times504$, \\
     & $768\times432$, $640\times360$, $512\times288$, \\
     & $384\times216$, $256\times144$, $128\times72$ \\
    Request rate & 30 Hz \\
    Individual eval. time & 10 minutes \\
    \bottomrule
  \end{tabular}
\end{table}

We evaluate five protocols using the MAR system over 5G, utilising open-source protocol libraries of UDP\footnote{In-built Python \lstinline{socket} library}, TCP\footnotemark[1], MPEG-TS\footnote{https://github.com/kkroening/ffmpeg-python}, RTP\footnote{https://gitlab.com/nickvsnetworking/pyrtp}, and QUIC\footnote{https://github.com/aiortc/aioquic}. Table \ref{tab:experiment_parameters} presents a summary of the parameters used in the evaluation. In addition to varying the protocol, the other varying parameter is the resolution of the images before they are sent from the client to server, as this, in effect, evaluates different transmission rates.

We maintain a 16:9 aspect ratio for each of the image resolutions as the video codecs function best when utilising width and height dimensions which are multiples of 16, 8, and 4 \cite{brightcove:videodimensions}. Specifically, we use a resolution range which are multiples of 8 as this produces a good spread for testing. We also note that while not explicitly evaluating video transmission, our capture frame rate is 30 Hz, which approximates to standard 30 FPS video. 

When collecting data, we configure the client application to make requests for a ten minute evaluation period as this allows for a good indication of the protocol performance under active network conditions while averaging out noise or momentary changes in network quality.

For collecting performance data, we utilise the passive network measurement tool Qosium \cite{kaitotek:qosium}, which captures and logs traffic. A Qosium software probe is attached to both the OnePlus smartphone and the edge PC. The data is collected by a corresponding Qosium scope run as another process on the edge PC. To ensure that we obtain accurate results from Qosium, we synchronise the client and server OS clocks using the master-slave precision time protocol. Additionally, the smartphone is kept fully charged to maintain constant system performance and we disable background processes.

\section{Results and Analytics} \label{sec:analysis}
\begin{figure}
    \begin{center}
      \centering
      \includegraphics[width=\linewidth]{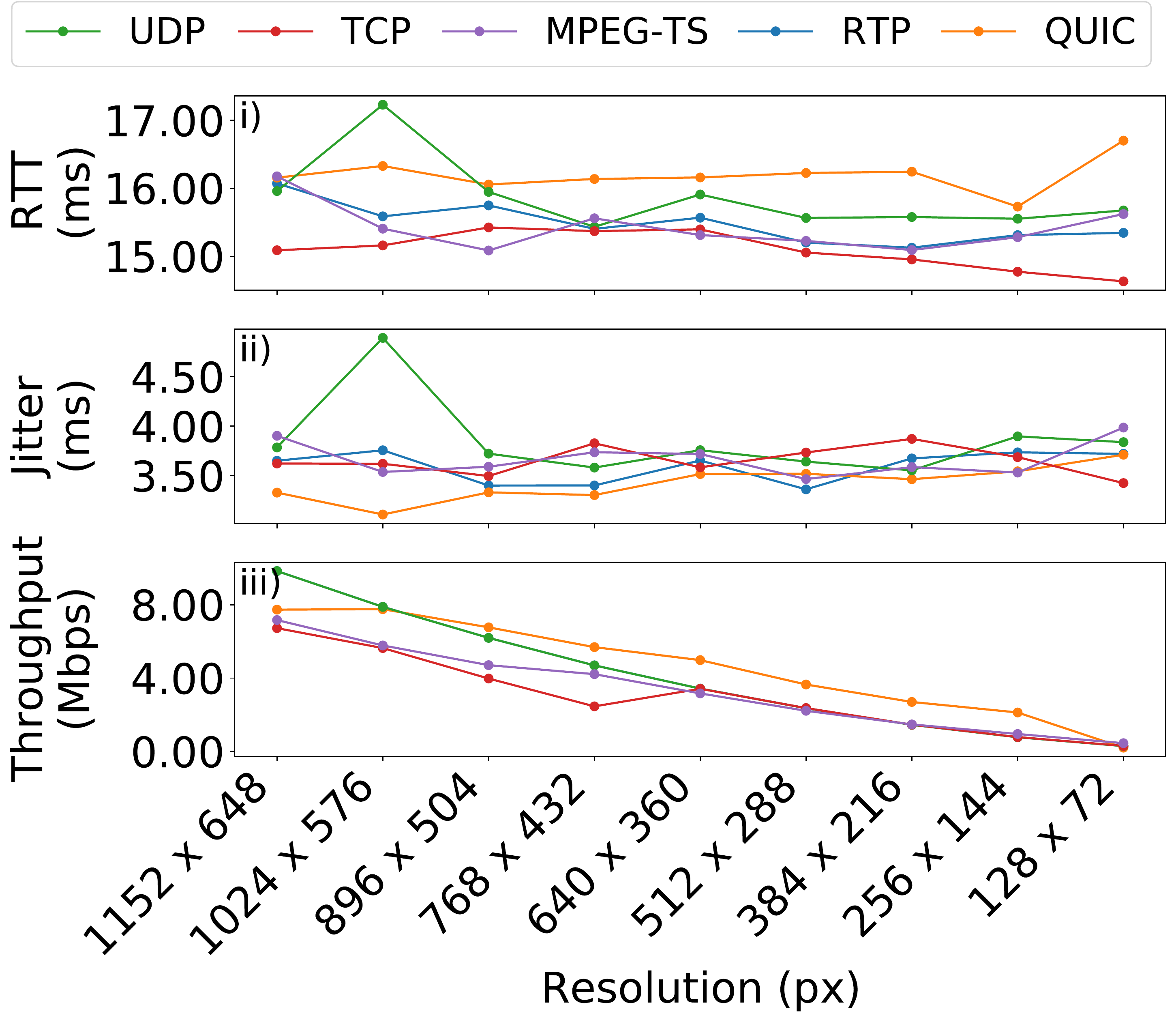}
      \caption{Evaluation results for different transport protocols and resolution sizes with 5G testbed.}
      \label{fig:complete_5g}
    \end{center}
\end{figure}

\begin{figure}
    \begin{center}
      \centering
      \includegraphics[width=\linewidth]{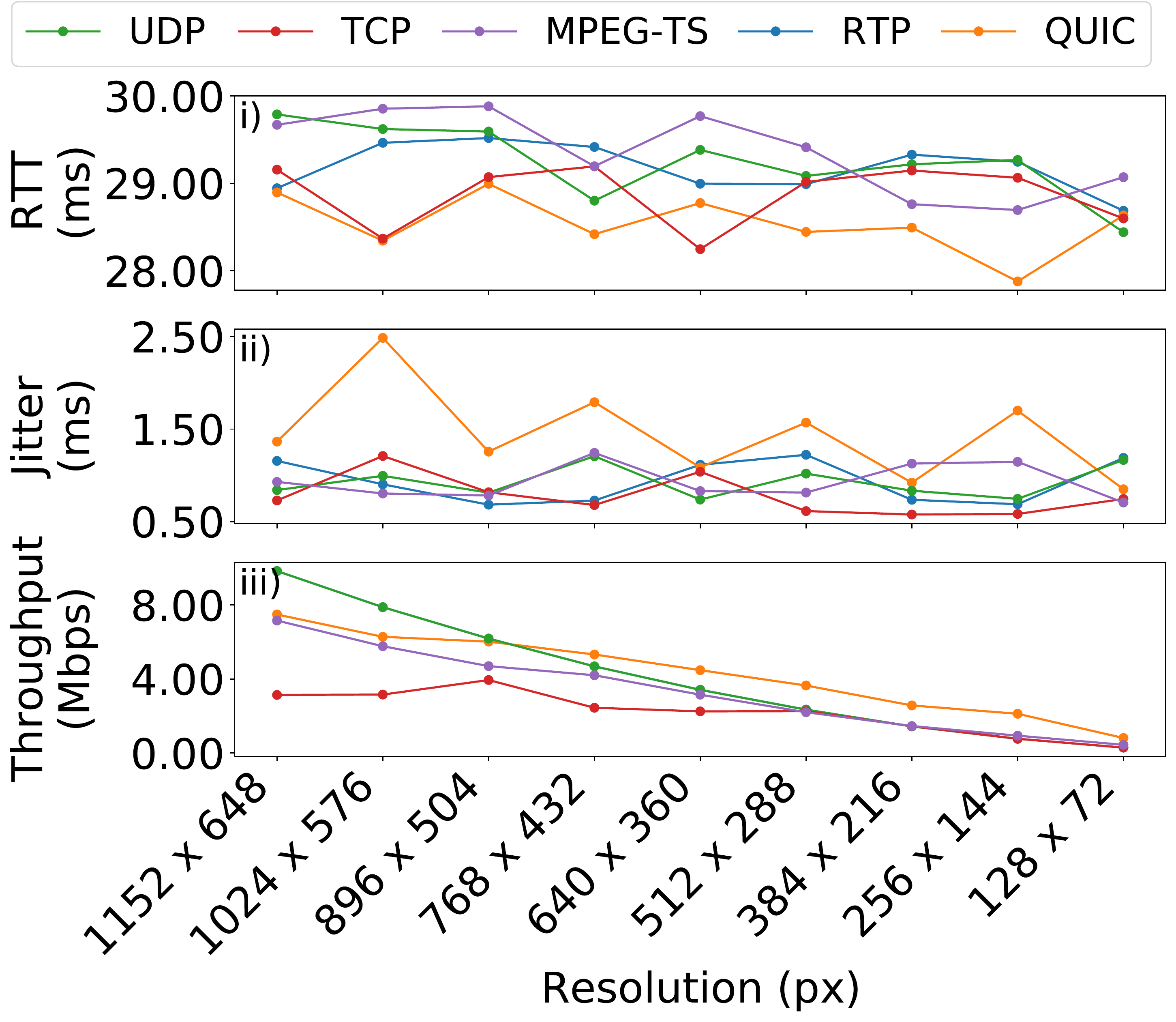}
      \caption{Evaluation results for different transport protocols and resolution sizes while connected to LTE.}
      \label{fig:complete_4g}
    \end{center}
\end{figure}

\begin{figure}
    \begin{center}
      \centering
      \includegraphics[width=\linewidth]{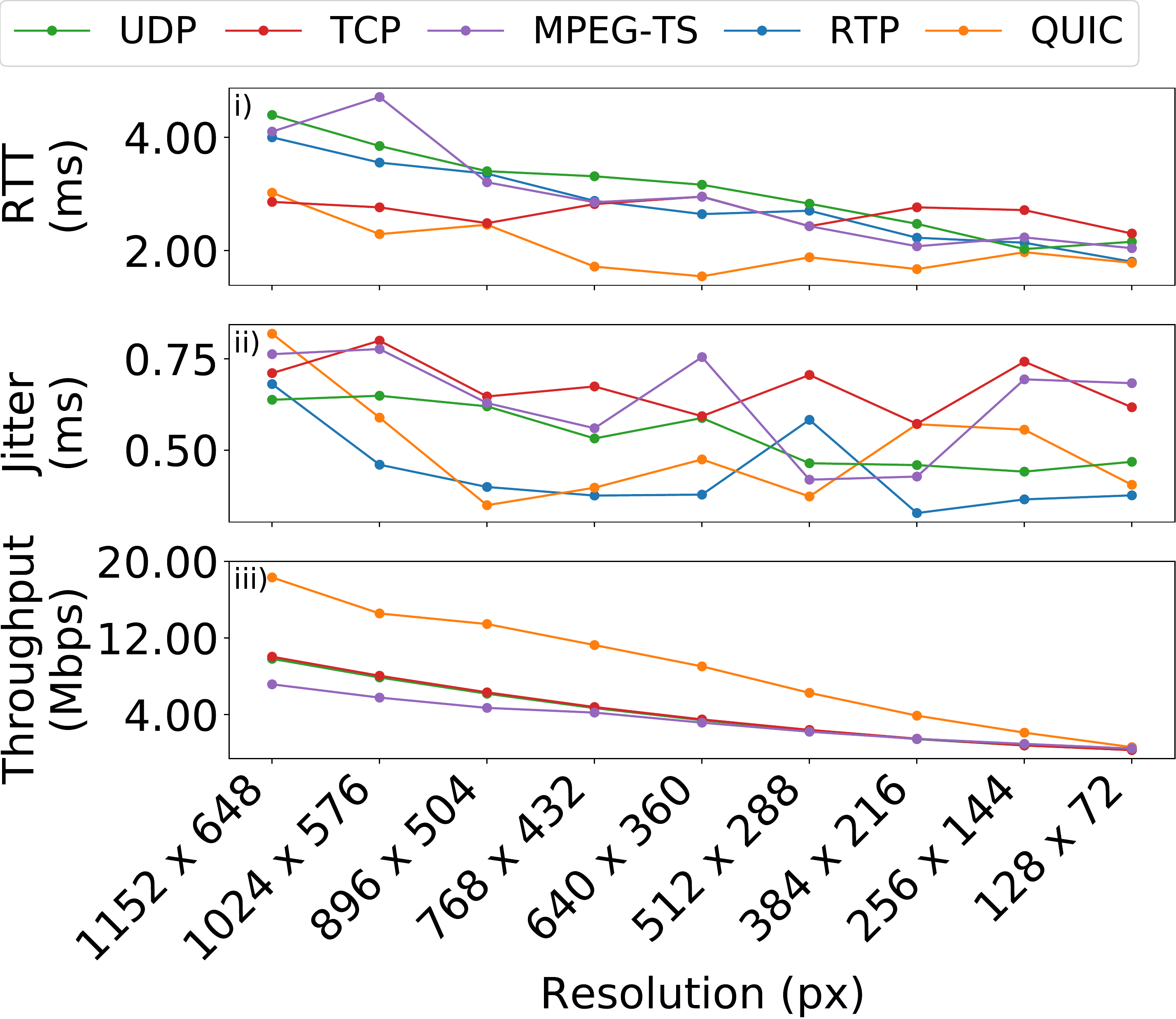}
      \caption{Evaluation results for different transport protocols and resolution sizes while connected to WiFi.}
      \label{fig:complete_wifi}
    \end{center}
\end{figure}

Figure \ref{fig:complete_5g} presents evaluation results of the different protocols on the 5G testbed given metrics of RTT (the time between client sending a request and receiving a reply), jitter (variation in packet delay from client to server) \cite{ferrari1992delay}), and throughput (data transferred per second from client to server over the ten minute test). 
Furthermore, Figures \ref{fig:complete_4g} and \ref{fig:complete_wifi} present corresponding evaluation results on LTE and WiFi. 



\textbf{RTT in 5G vs LTE \& WiFi.} For 5G, the median and standard deviation RTT (Figure \ref{fig:complete_5g}(i)) across all protocols is $15.56\pm0.51$ ms. Except for QUIC, the RTT of all protocols decreases as image resolution decreases. From the largest to the smallest resolution, the average RTT decrease is $\approx3.03\%$. Amongst the five protocols, TCP has the lowest RTT for seven different resolutions and has a median of $15.09\pm0.26$ ms. In comparison, QUIC has the highest median 5G RTT of $16.16\pm0.24$ ms, the corresponding LTE (Figure \ref{fig:complete_4g}(i)) and WiFi (Figure \ref{fig:complete_wifi}(i)) results show that, with these connections, QUIC has the smallest medians of $28.49\pm0.32$ ms and $1.88\pm0.44$ ms, respectively. 

\begin{table}
  \caption{QUIC vs. TCP results with good and degraded signal quality for large and small resolution sizes on 5G}
  \label{tab:degraded_connection}
  \begin{tabular}{cccc}
    \toprule
    Protocol & Location & Resolution & RTT (ms) \\
    \midrule
    QUIC & Main & $1152\times648$ & 16.37 \\
     & Main & $128\times72$ & 15.79 \\
     & Second & $1152\times648$ & 16.18 \\
     & Second & $128\times72$ & 15.96 \\
    TCP & Main & $1152\times648$ & 15.27 \\
     & Main & $128\times72$ & 14.73 \\
     & Second & $1152\times648$ & 15.885 \\
     & Second & $128\times72$ & 15.23 \\
    \bottomrule
  \end{tabular}
\end{table}

\begin{figure*}
  \begin{subfigure}{0.49\textwidth}
    \includegraphics[width=\linewidth]{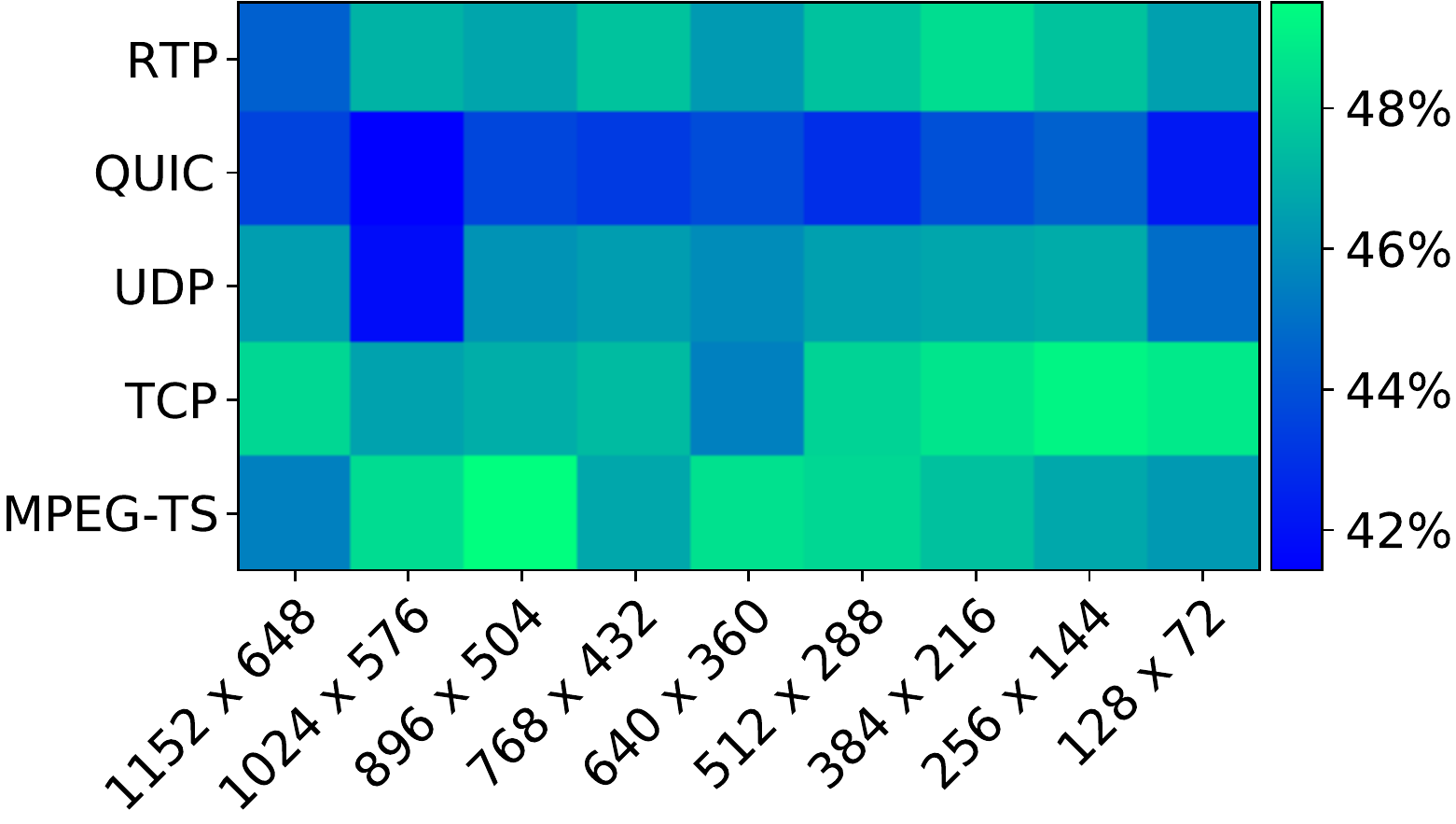}
    \caption{5G vs. LTE} 
  \end{subfigure}%
  \begin{subfigure}{0.49\textwidth}
    \includegraphics[width=\linewidth]{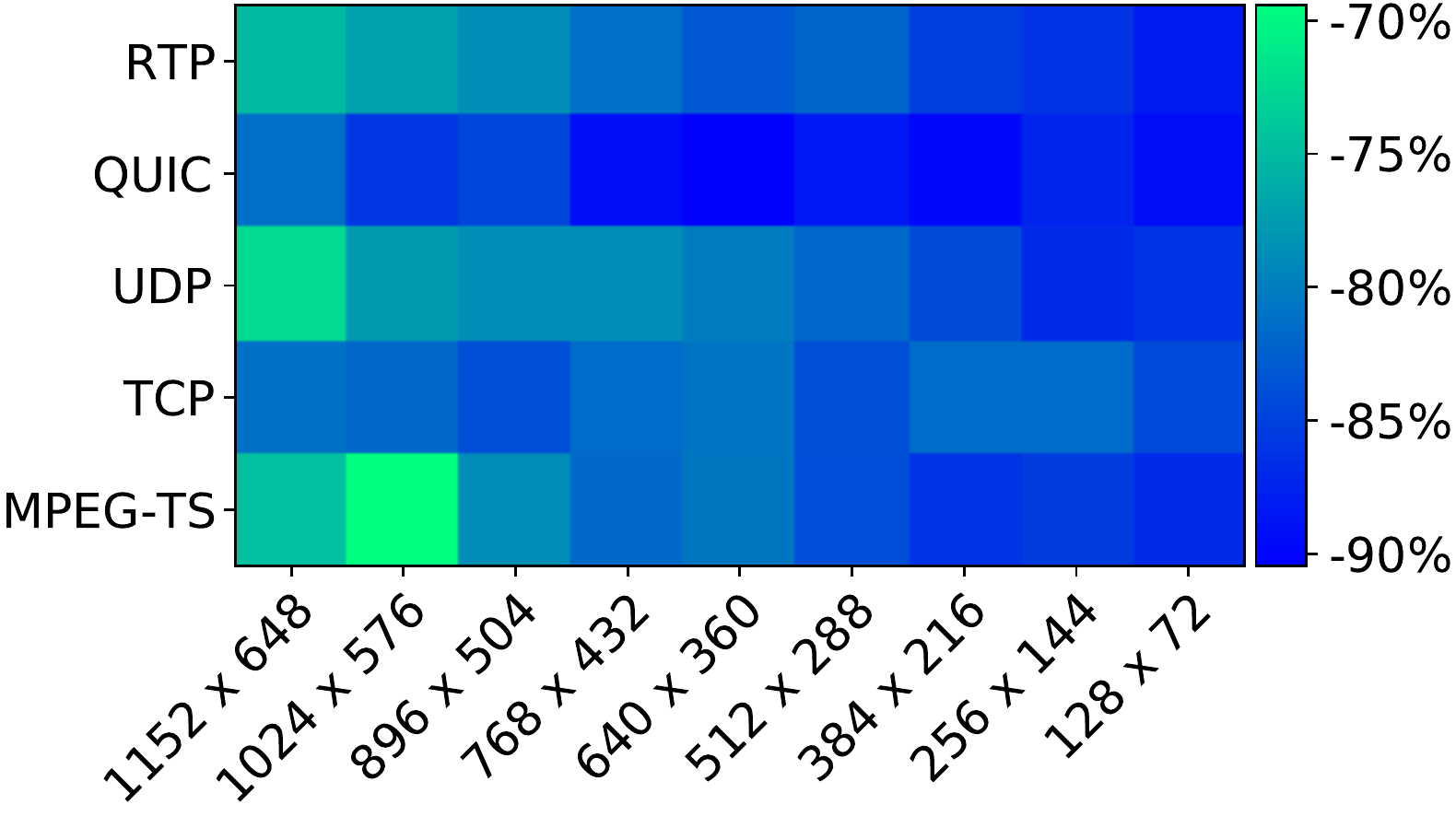}
    \caption{5G vs. WiFi} 
  \end{subfigure}%
  \caption{Heatmaps of the percentage difference, showing improvement or deterioration in RTT performance for the protocols at each resolution for the three network types: (a) 5G vs. LTE, and (b) 5G vs. WiFi} 
  \label{fig:heatmaps_rtt}
\end{figure*}

Given this difference in relative performance for QUIC between 5G and LTE/WiFi, we further evaluate QUIC and TCP on 5G under the same conditions as previously, i.e., 30 Hz request rate for a total of ten minutes, except limited to the two resolution extremes ($1152\times648$ with file size 151.2kB, and $128\times72$ at 4.5 kB), and we vary the signal quality to the 5G network as denoted by the 5G metric, SS-RSRQ. The main test location has an SS-RSRQ of $\approx-6$ dB, denoted as good signal quality, and the secondary location has a deteriorated SS-RSRQ of $\approx-10$ dB, denoted degraded signal quality.

Table \ref{tab:degraded_connection} shows that for QUIC, the $1152\times648$ resolution RTT decreases by 0.19 ms between the main and secondary testing locations, but increases by 0.17 ms for $128\times72$. while for TCP, the RTT increases for the higher resolution in both locations. This suggests that QUIC improves protocol performance when larger object sizes are transferred and when signal quality is degraded, e.g., under LTE network conditions relative to 5G; as we initially suspected. Other studies such as \cite{megyesi2016quick}, \cite{biswal2016does}, and \cite{yu2017quic} also support these findings. They postulate that lower connection establishment latency and improved congestion control mechanisms are the reasons for this behaviour. 


Comparisons of the protocols RTT performance between 5G, LTE, and WiFi are presented as percent difference heatmaps in Figure \ref{fig:heatmaps_rtt}. We observe from Figure \ref{fig:heatmaps_rtt}(a) that, from LTE to 5G, QUIC shows the smallest RTT improvement of just $43.2\%$, while TCP shows the largest improvement of $48.1\%$. Overall, 5G provides a median $47\%$ improvement in RTT over LTE. However, the WiFi connection still outperforms 5G by $82\%$. This is attributed to the close geographical location of the WiFi hotspot and because only two devices are connected to the hotspot, whereas the mobile network is a live network with substantially more devices connected.  

\textbf{Jitter in 5G vs LTE \& WiFi.} From Figure \ref{fig:complete_5g}(ii), we observe that jitter decreases as image resolution decreases for different protocols on 5G, however, there are anomalies. QUIC and RTP have the smallest jitter for 5G and WiFi, while TCP has the smallest for LTE. Also for 5G, UDP has a spike at the $1024\times576$ resolution. Overall, the decreasing trend is not consistent throughout (e.g., UDP's spike), and while decreasing in WiFi (Figure \ref{fig:complete_wifi}(ii))), jitter appears to increase with resolution in LTE (Figure \ref{fig:complete_4g}(ii)). The jitter inconsistency could be related to mobile network congestion and the resulting instability due to traffic from a large number of 4G users and devices. The differing jitter ranges, i.e., 3.11-4.81 ms for 5G, 0.58-2.48 ms for LTE, and 0.33-0.82 ms for WiFi, could also be a result of the differing frequencies and physical distances between the client and APs, namely, the LTE signal has a lower frequency than 5G (2600 MHz vs. 3500 MHz), and the LTE Picocell and WiFi AP are closer to the 5G smartphone.


\textbf{Throughput in 5G vs LTE \& WiFi.} All three network types (5G, LTE, and WiFi) show similar throughput as seen in Figures \ref{fig:complete_5g}(iii), \ref{fig:complete_4g}(iii), and \ref{fig:complete_wifi}(iii). The trends are also similar (and show a decrease) and the similarity is especially notable for the throughputs of RTP and UDP, which are within a mean standard deviation of 0.003 Mbps of each other. Whereas, QUIC provides on average the highest throughput, i.e., $17.5\%$ higher than UDP with a 5G connection and $17.9\%$ with LTE. Surprisingly, QUIC has a stronger advantage on WiFi, with a $57.4\%$ higher throughput than UDP. The throughput of TCP when using an LTE connection stands out in particular, as the throughputs for the resolutions $1152\times648$ and $1024\times576$ are $53.3\%$ lower than their counterparts collected with 5G.

\begin{table*}
  \caption{Comparison of the general requirements of a MAR protocol with the features of QUIC, RTP, and TCP.}
  \label{tab:technical_comparison}
  \begin{tabular}{p{2.5cm}p{4.5cm}p{4.5cm}p{4.5cm}}
    \toprule
    MAR Protocol Requirements \cite{braud2017future} & QUIC \cite{hamilton2016quic, langley2017quic, thomson2019using} & RTP \cite{ietf2003rtp, zopf2002real} & TCP \cite{postel1981transmission} \\
    \midrule
    Designed purposes & Web apps and video streaming & Video and audio transmission & Agnostic to data type \\
    \hline
    Low-latency and high fault tolerance & Low-latency through combining protocol version negotiation with other handshakes (cryptographic and transport). Fault tolerance through ack-based retransmitting of lost packets & No mechanism for low latency. No in-built fault tolerance to ensure packet delivery, nor for preventing out-of-order delivery. & Fault tolerance through retransmitting lost segments and duplicate segments are discarded by using sequence numbers and acknowledgement packets.  \\
    \hline
    Fair to other concurrent network connections & Minimises per-packet bandwidth and computational costs by packaging many frames within a single packet & Reduce bandwidth with discontinuous transmission (silence suppression), i.e., not transmitting audio data & Unfairness from buffer availability, separate implementations could prevent this \cite{pilosof2003understanding, leith2005tcp} \\
    \hline
    Multipath transmission & No built-in multipath, draft multipath QUIC extensions allow  & No built-in multipath, draft multipath RTP extensions allow & No built-in multipath, standard MPTCP \cite{ford2011architectural} extends TCP for multipath. \\
    \bottomrule
  \end{tabular}
\end{table*}

\textbf{Takeaway.} From the 5G-based evaluation results, the key takeaways are as follows. TCP is the best performing protocol by RTT. Though our SS-RSRQ test suggests QUIC has better performance when signal quality is degraded and QUIC's throughput out-performs that of other protocols. We additionally compare the general requirements for a MAR protocol with the features of TCP (the best protocol by RTT), QUIC (shows promise), and RTP (a multimedia-focused protocol). Summarised in Table \ref{tab:technical_comparison}, we see that while QUIC and TCP have in-built fault tolerance through retransmission mechanisms, guaranteed low latency will require the system addition of 5G edge. While each protocol minimises their bandwidth costs, 5G networks dynamically optimise connection bandwidths according to usage and demand.

\section{Conclusions} \label{sec:conclusions}
We evaluated the performance of five transport protocols using a MAR system over a 5G testbed with metrics including RTT, jitter, and throughput. The evaluation also compared their performances against LTE and WiFi. We also compare the general MAR protocol requirements with the technical features of three protocols (the two with the best evaluation results, QUIC and TCP, and a protocol intended for multimedia transmission, RTP). From our quantitative analysis of 5G results, TCP is the most promising candidate as a MAR transport protocol in 5G. By comparison, an analysis of the LTE results indicate that QUIC performs better on LTE. Additionally, we preliminary investigate the impact on protocol performance of varying signal quality. 


As our current evaluations are limited to only a few tested contexts, such as the signal quality, frequencies, and protocols, in future work we will further quantify the impact of network quality on protocol performance with 5G and LTE mobile connections. In addition, we will evaluate alternative transport protocols such as SCTP and MPTCP, and test with scalable scenarios, i.e., with multiple servers and clients. Finally, we will develop and utilise QoE-derived metrics to evaluate protocol performance for MAR, which will complement the network-based metric results.
\bibliographystyle{ACM-Reference-Format}
\bibliography{bibliography}

\end{document}